\documentstyle[11pt]{article}

\def\ket#1{| #1 \rangle}

\def\BraKet#1#2{\langle #1 | #2 \rangle}

\begin{document}

\centerline{\Large\bf On the Distinguishability of Relativistic Quantum States} 
\vskip 2mm
\centerline{\Large\bf in Quantum Cryptography} 
\vskip 3mm
\centerline{R.Laiho$^{\dag}$, S.N.Molotkov$^{\ddag}$, and S.S.Nazin$^{\ddag}$} 
\centerline{\it\small $^{\dag}$Wihuri Physical Laboratory,} 
\centerline{\it\small Physical Department, University of Turku, 20014 Turku, Finland} 
\vskip 2mm
\centerline{\sl\small $^{\ddag}$Institute of Solid State Physics of 
                      Russian Academy of Sciences,}
\centerline{\sl\small Chernogolovka, Moscow District, 142432 Russia}
\vskip 3mm

\begin{abstract}
Relativistic quantum field theory imposes additional fundamental
restrictions on the distinguishability of quantum states. Because of the
unavoidable delocalization of the quantum field states in the Minkowski
space-time, the reliable (with unit probability) distinguishability
of orthogonal states formally requires infinite time. For the cryptographic 
protocols which are finite in time the latter means that the effective 
``noise'' is present even in the ideal communication channel because of the
non-localizability of the quantum field states.
\end{abstract}

\section{Introduction}
Employment of quantum states as the information carriers opened
new prospects for the realization of various cryptographic protocols [1--5].
An important result in this field was achieved when it was shown
that within the framework of the non-relativistic quantum mechanics
some basic classical cryptographic protocols (whose security in classical
physics is based on the computational complexity only) become unconditionally
secure, i.e. their security is based on the laws of nature (non-relativistic
quantum mechanics) and cease to depend on the available computational 
resources. However, for a number of basic tasks, e.g. bit commitment,
all the numerous efforts to develop an unconditionally secure quantum protocol
failed [6--9]. Moreover, it was shown later that no ideal unconditionally 
secure bit commitment protocol can be constructed in the framework of the
non-relativistic quantum mechanics [10-11].

The only information carriers suitable for the realistic large-scale
quantum cryptographic systems are the photons which are essentially
relativistic particles. Since the relativistic quantum mechanics does not
allow any sensible physical interpretation, the theory describing the
relativistic quantum systems arises as already the quantum field theory [12].
The relativistic quantum field theory imposes additional fundamental
restrictions on the speed of information transfer and the processes
of the quantum field state measurements. The latter circumstance has the
consequence that the secret protocols which cannot be realized in the
non-relativistic quantum mechanics become realizable in the classical
relativistic theory and in the quantum field theory [13--15].

Below we shall show that in the relativistic quantum field theory
the reliable (with unit probability) distinguishability of two orthogonal 
states requires an infinite time (which actually is a consequence of
the microcausality principle), in contrast to the non-relativistic
quantum mechanics where this can be done instantly at any time.
This circumstance allows to construct new cryptographic
protocols in the quantum field theory.

All quantum cryptographic protocols actually employ the following two
features of quantum theory. The first one is the no cloning theorem [16], 
i.e. the impossibility
of copying of an arbitrary quantum state which is not known beforehand or,
in other words, the impossibility of the following process:
\begin{displaymath}
  \ket{A}\ket{\psi} \rightarrow U(\ket{A}\ket{\psi})=
                 \ket{B_{\psi}}\ket{\psi}\ket{\psi},
\end{displaymath}
where $\ket{A}$ and $\ket{B_{\psi}}$ are the
apparatus states before and copying act, respectively, and
$U$ is a unitary operator. Such a process is prohibited
by the linearity and unitary nature of quantum evolution.
Actually, even a weaker process of obtaining any information about
one of the two non-orthogonal states without disturbing it is impossible,
i.e. the final states of the apparatus $\ket{A_{\psi_1}}$ 
and $\ket{A_{\psi_2}}$ corresponding to the initial input states
$\ket{\psi_1}$ and $\ket{\psi_2}$, respectively, after the unitary 
evolution $U$, 
\begin{displaymath}
  \ket{A}\ket{\psi_1} \rightarrow U(\ket{A}\ket{\psi_1})=
                         \ket{A_{\psi_1}}\ket{\psi_1} ,
\end{displaymath}
\begin{displaymath}
  \ket{A}\ket{\psi_2} \rightarrow U(\ket{A}\ket{\psi_2})=
                         \ket{A_{\psi_2}}\ket{\psi_2} ,
\end{displaymath}
can only be different, $\ket{A_{\psi_1}} \neq \ket{A_{\psi_2}}$, if 
$\BraKet{\psi_1}{\psi_2} = 0$ [10], which means the impossibility 
of reliable distinguishing between non-orthogonal states.
There is no such a restriction for orthogonal states.
Moreover, in the non-relativistic quantum mechanics generally there is no
restriction on the instant (arbitrarily fast) reliable distinguishability 
of any two orthogonal states at any time without disturbing them. 

That is why the orthogonal states are not even discussed in the
context of non-relativistic quantum cryptographic protocols.

On the contrary, in the quantum field theory it turns out that the
distinguishability of two orthogonal states with the probability arbitrarily
close to unity requires finite time, and the reliable distinguishability
(with unit probability) can only be achieved in infinite time.

To clarify the difference between the non-relativistic quantum mechanics
and the relativistic quantum field theory, consider first the non-relativistic
case.

Suppose we have a pair of orthogonal states in the Hilbert space:
$|\psi_{1,2}\rangle\in {\cal H}$ and $\langle\psi_1|\psi_2\rangle=0$. 
To reliably distinguish these states, one can for example use the following 
orthogonal identity resolution in ${\cal H}$:
\begin{equation}
{\cal P}_1+{\cal P}_2+{\cal P}_{\bot}=I,\quad 
{\cal P}_{1,2}=|\psi_{1,2}\rangle\langle\psi_{1,2}|, \quad 
{\cal P}_{\bot}=I-{\cal P}_1-{\cal P}_2, 
\end{equation} 
where ${\cal P}_{1,2}$ are the projectors on the subspaces 
${\cal H}_{1,2}$ spanned by the states $|\psi_1\rangle$ and $|\psi_2\rangle$,  
while ${\cal P}_{\bot}$ is the projector on
${\cal H}_{12}^{\bot}=({\cal H}_1\oplus{\cal H}_2)^{\bot}$.  
The probability of obtaining, for example, the result in the channel
${\cal P}_1$ if in input state was $|\psi_1\rangle$, is 
\begin{equation} 
\mbox{Pr}_1\{|\psi_1\rangle\}=
\mbox{Tr}\{|\psi_1\rangle\langle\psi_1|{\cal P}_1\}
\equiv 1,
\end{equation}
while it is identically equal to zero in the channels ${\cal P}_{2,\bot}$:
\begin{equation}
\mbox{Pr}_{2,\bot}\{|\psi_1\rangle\}=
\mbox{Tr}\{|\psi_1\rangle\langle\psi_1|{\cal P}_{2,\bot}\}
\equiv 0,
\end{equation}
and similarly for the input state $|\psi_2\rangle$. 
The equations (1--3) mean that the orthogonal states can be reliably
(with unit probability) distinguished. To answer the question of whether or
not this can be done instantly and without disturbing the measured states,
one should consider the measurement procedure in more detail. Indeed,
there are three natural levels of description of the measurement process
in quantum mechanics which differ in the amount of information they provide
[17--19].  The simplest description of the measurement procedure lists
only the possible measurement outcomes (i.e., specifies the space of possible
measurement outcomes) and specifies the relative frequencies (probabilities)
of occurrence of a particular outcome for a given input state of the
measured quantum system. In that sense the measurements are in one-to-one 
correspondence with the positive identity resolutions in the Hilbert
space of the system states [17--19]. However, this approach completely
ignores the problem of finding the state of the system after the
measurement which gave a particular result. At the next level of description
of quantum mechanical measurement each measurement procedure is associated
with the so-called instrument $T$ which actually is, speaking informally,
a rule allowing to ascribe to each pair consisting of the input density 
matrix $\rho_0$ of the measured system and the measurement result $i$ 
a positive matrix $\tilde\rho_i$ with the trace $\rm{Tr}\tilde\rho_i=p_i< 1$, 
which is interpreted as an unnormalized density matrix of the ensemble
of the systems selected by the condition that the performed measurement 
procedure gave the outcome $i$ (the probability of obtaining the $i$-th 
outcome being $p_i$ and the state of the system after the outcome $i$
is obtained being $\rho_i=\tilde\rho_i/p_i$). In this approach the
von Neumann -- L\"uders projection postulate is equivalent to
the statement that the instrument $T$ corresponding to the measurement
described by Eqs.(1--3) is given by the formula
\begin{displaymath}
\tilde{\rho}_i=
{\cal P}_i \rho {\cal P}_i,
\end{displaymath}
so that for example for the input state $|\psi_1\rangle$ we have
\begin{equation}
\tilde{\rho}_1=
\frac{ {\cal P}_1|\psi_1\rangle\langle\psi_1|{\cal P}_1}
{ \mbox{Tr} \{ |\psi_1\rangle\langle\psi_1|{\cal P}_1 \} }=
|\psi_1\rangle\langle\psi_1|.
\end{equation}
Equation (4) means that under the assumption of the possibility
of the realization of von Neumann measurement satisfying the projection
postulate the orthogonal states can be reliably distinguished without
any disturbance. One should also note that the reliable distinguishing
of orthogonal states without any disturbance can be performed at arbitrarily
chosen moment of time, i.e. the measurement procedure can be started at any
moment although the procedure itself generally depends on the chosen 
starting moment. Indeed, since the states $|\psi_1\rangle$ and 
$|\psi_2\rangle$ are assumed to be known and one should only determine
which of these two states is actually given for a particular measurement
run, the temporal evolution of these states ($|\psi_i(t)\rangle=U(t)|\psi_i\rangle$, 
where $|\psi_i\rangle$ is the system state at $t=0$) 
should also be considered as a known function of time. If the measurement
is performed at time $t$, the identity resolution should employ
${\cal P}_i(t)=U(t){\cal P}_iU^{-1}(t)$ as the corresponding projectors.

An important point is that the duration of the measurement procedure
$\tau$ has not yet been actually mentioned. To introduce $\tau$, one
should consider the third, most detailed, level description
of the measurement [18,19]. In the quantum mechanical measurement theory 
it is shown that any instrument can be represented in the following way:
first the measured quantum system $S$ begins to interact with
an auxiliary system $A$ (ancilla) and they perform a joint
evolution during time $\tau$. Then the system $A$ is subjected to 
a von Neumann measurement (assumed to be completed arbitrarily fast)
whose outcome determines the state of system $S$ immediately after 
the measurement. Therefore, the speed with which a particular measurement
procedure can actually be realized depends on the possibility of choosing a 
suitable auxiliary system $A$ and the realization of required interaction
between $A$ and $S$.

\section{Non-relativistic case}
Since the statement of the possibility of reliable distinguishability at any 
time of the two orthogonal states is quite general, we shall make
the transition to the relativistic case smoother by considering 
in the non-relativistic case two states $|\psi_1\rangle$ and $|\psi_2\rangle$
of a one-dimensional free particle.

It is intuitively clear that to identify a state one should have access to
the entire spatial domain where the wavefunction of the particle is present.
In the non-relativistic case there are no restrictions on the maximum speed
of the information transfer so that an entire extended spatial domain can be 
accessed instantly at any time, i.e. there are no restrictions on instant
non-local measurements.

It is natural to assume that the identification of the state of a quantum field
also requires access to the entire domain of the Minkowski space-time where the 
field is present. However, the existence of the maximum speed of information 
transfer no extended domain can be accessed instantly, i.e. non-local any 
measurement requires a finite time.

In the non-relativistic quantum mechanics the above considerations are obvious
because the states of a non-relativistic particle can be described by a 
{\it wavefunction}. In the relativistic quantum field theory these
intuitively appealing considerations should be analyzed more rigorously, 
since the states of a quantum filed cannot be described by a {\it wavefunction} 
(or even by an operator field).

The set of observables of a quantum system is represented by
an operator algebra with a unit and a conjugation (involution) which are
defined on a dense subspace $\Omega\subset {\cal H}$ so that 
$A|\varphi\rangle\in\Omega$ if $|\varphi\rangle\in\Omega$. 
The Hilbert state ${\cal H}$ itself is a completion of $\Omega$ 
with respect to the convergence defined by the norm induced by the scalar 
product in ${\cal H}$. The convergence conditions in $\Omega$ are stronger 
than in ${\cal H}$ and can be chosen in a suitable way for a particular
physical system. Continuous functionals defined on elements
$|\varphi\rangle\in\Omega\subset{\cal H}$
constitute a linear space $\Omega^*$ dual to $\Omega$. 
The space $\Omega^*$ contains more elements than 
${\cal H}^*$ dual to ${\cal H}$. The space 
${\cal H}^*$ consists of continuous linear functionals defined on 
elements $|\varphi\rangle\in{\cal H}$, frequently written as
$\langle f|\in {\cal H}^*$ (the value of a functional is a $c$-number 
$\langle f|\varphi\rangle$). The space ${\cal H}^*$ is actually known to
be isomorphic to ${\cal H}$ itself.

The elements of $\Omega^*$ frequently arise as the generalized eigenvectors
$|\psi_{\alpha}\rangle$ of the operator $A$ with a continuous spectrum
\begin{equation}
A|\psi_{\alpha}\rangle=\lambda_{\alpha}|\psi_{\alpha}\rangle,
\quad |\psi_{\alpha}\rangle\in\Omega^*,
\end{equation}
\begin{displaymath}
\langle\psi_{\alpha}|\psi_{\alpha'}\rangle=\delta(\alpha-\alpha').
\end{displaymath}
Any state vector 
$|\varphi\rangle\in\Omega\subset{\cal H}$
can be expanded in a complete set of generalized eigenvectors
\begin{equation}
|\varphi\rangle=\int \varphi(\alpha) |\psi_{\alpha}\rangle d\alpha,
\end{equation}
where the coefficients (wavefunctions) $\varphi(\alpha)$ are given 
by the values of a functional from $\Omega^*$
\begin{equation}
\varphi(\alpha)=\langle\psi_{\alpha}|\varphi\rangle.
\end{equation}
The construction $\Omega\subset{\cal H}\subset\Omega^*$ is called 
rigged Hilbert space (Gel'fand triplet) [22--24].

The self-adjoint operator with a continuous spectrum possesses a spectral
resolution built of the spectral projectors associated with
the generalized eigenvectors
\begin{equation}
A=\int \lambda_{\alpha}|\psi_{\alpha}\rangle\langle\psi_{\alpha}|
d\alpha, 
\end{equation} 
and the corresponding identity resolution 
\begin{equation} 
I=\int |\psi_{\alpha}\rangle\langle\psi_{\alpha}|d\alpha.  
\end{equation}
A particular functional realization of the rigged space can be chosen
in the form of ${\cal J}(x)\subset {\cal L}_2(x,dx)\subset {\cal J}^*(x)$
(where ${\cal J}(x)$ is the space of smooth rapidly decreasing functions, 
${\cal L}_2(x,dx)$ is the space of square integrable functions, and
${\cal J}^*(x)$ is the space of tempered distributions [23--25]).

Consider two orthogonal states of a free non-relativistic one-dimensional 
particle: 
\begin{equation}               
|\varphi_{1,2}\rangle=\int_{-\infty}^{\infty}\varphi_{1,2}(x)|x\rangle dx,
\quad |\varphi_{1,2}\rangle\in {\cal J}(x)\subset{\cal L}_2(x,dx),
\end{equation}               
\begin{displaymath}
\langle\varphi_1|\varphi_2\rangle=0,\quad
|x\rangle\in {\cal J}^*(x),\quad \langle x|x'\rangle=\delta(x-x'),
\end{displaymath}
where $|x\rangle$ is the generalized eigenvector of the position operator. 
The identity resolution based on the generalized eigenvectors is familiarly 
written as
\begin{equation}               
I= \int_{-\infty}^{\infty} |x\rangle\langle x| dx.
\end{equation}               
The measurement allowing to reliably distinguish between the two states
$|\varphi_1\rangle$ and $|\varphi_2\rangle$  
without disturbing them is given by the identity resolution defined by Eq.(1).
For the measurement outcomes we have
\begin{equation}               
\mbox{Pr}_i\{|\varphi_j\rangle\}=\mbox{Tr}
\{ |\varphi_j\rangle\langle\varphi_j| P_j\}=
\mid\int_{-\infty}^{\infty}\int_{-\infty}^{\infty}
\varphi^*_j(x)\delta(x-x')\varphi_i(x')dxdx'\mid^2=\delta_{ij}.
\end{equation}
This measurement is non-local in the sense that the reliable state
identification requires the access to the entire spatial domain where
the wavefunction is different from zero. In the non-relativistic
quantum mechanics there is no restriction on the maximum speed of
information transfer so that the non-local measurements can be
made instantly (arbitrarily fast).

Thus, the physical apparatus implementing the measurement (1--3,12)
should be non-local in space. The spatial position of the observer which is 
the ultimate element in any measurement procedure can be chosen
completely arbitrarily since due to the unlimited speed of information 
transfer the data read off with any non-local device can be 
instantly gathered to any spatial point where the observed is located.

\section{Relativistic case}
Consider now the relativistic case. The states of a relativistic quantum system
(field) are described by the rays in the physical Hilbert space ${\cal H}$
where a unitary representation of the Poincar\'e group is realized [23,24].
The local quantum field $\varphi(\hat{x})$ (here $\hat{x}=(t,{\bf x})$ is
a point in the Minkowski space) is defined as a tensor (if the field
has more than one component) operator valued distribution. To be more precise,
any function (or a set of function if the field has several components)
$f(\hat{x})\in {\cal J}(\hat{x})$ is associated with an operator $\varphi(f)$ 
which can formally be written as 
\begin{equation}               
\varphi(f)=\sum_{j=1}^{r}\varphi_j(f_j)= 
\sum_{j=1}^{r}  \int\varphi_j(\hat{x})f_j(\hat{x})d\hat{x},
\end{equation}               
All the operators $\varphi(f)$ and $\varphi^*(f)$ have a common domain 
$\Omega$ which does not depend on $f(\hat{x})$ and is a dense subspace 
${\cal H}$ mapped by the operators into itself,
$\varphi(f)\Omega\subset\Omega$ ($\varphi^*(f)\Omega\subset\Omega$),
so that for the vectors $|\phi\rangle,{} |\psi\rangle\in \Omega\subset {\cal H}$,
the quantity $\langle\phi|\varphi(f)|\psi\rangle$ is a distribution 
from ${\cal J}^*(\hat{x})$.

The subspace $\Omega$ contains a cyclic vector, called the vacuum state, 
$|0\rangle\in\Omega$, such that the set of all polynomials 
$P(\varphi,f)$ (field operator algebra) generate the entire $\Omega$. 
The field operator algebra elements are defined as
\begin{equation}               
P(\varphi,f)=f_0+\sum_{n=1}^{\infty}\int\int\ldots\int
\varphi(\hat{x}_1)\varphi(\hat{x}_2)\ldots\varphi(\hat{x}_n)
f(\hat{x}_1,\hat{x}_2,\ldots, \hat{x}_n)d\hat{x}_1 d\hat{x}_2 \ldots 
d\hat{x}_n.
\end{equation}               
The field operators $\varphi(\hat{x})$ map the regular states from 
$\Omega$ to the generalized states $P(\varphi(\hat{x}))\Omega\subset\Omega^*$.
The microcausality principle is assumed to be satisfied or, to be more 
precise, if the functions $f(\hat{x}), g(\hat{y})$ have supports separated 
by a space-like interval 
($\mbox{supp}f(\hat{x})\cdot g(\hat{y})\in(\hat{x}-\hat{y})^2<0$), 
the field operators are assumed to either commute or anticommute, i.e. for any 
$|\psi\rangle\in\Omega$ the following microcausality relation holds:
\begin{equation}               
[\varphi(f),\varphi(g)]_{\pm}|\psi\rangle=0,\quad
(\hat{x}-\hat{y})^2<0.
\end{equation}     
At the infinity, the test functions from ${\cal J}(\hat{x})$ vanish faster
than any polynomial. However, the space of test functions ${\cal J}(\hat{x})$ 
contains a dense set of compactly supported functions 
${\cal D}(\hat{x})\subset {\cal J}(\hat{x})$ which are zero outside a
certain compact domain; therefore, any function from 
${\cal J}(\hat{x})$ can be approximated by a compactly supported function. 
The equation (15) should be interpreted as the statement
that for the field $\varphi(\hat{x})$ the measurements performed
in the domains separated by a space-like interval do not affect each other
since no interaction can propagate faster than light.

It is known [23,24] that if one requires that (i) the system states are described 
by the rays in a Hilbert space where a unitary representation of the Poincar\'e
group is realized and (ii) the spectrum of the group generators lies in
the front part of the light cone in the momentum representation then
the Lorentz-invariant quantum field can only be realized as an operator-valued
distribution rather than the field of operators $\varphi(\hat{x})$ acting in 
${\cal H}$. Therefore, one can only make sense of $\varphi(f)$
as an unbounded operator generating a state in $\Omega\subset{\cal H}$. 
The function $f(\hat{x})$ can be interpreted (with some reservations) 
as the amplitude (``shape'') of a one-particle packet.

The interpretation of a quantized field as a field of operators leads
to the trivial two-point function
$\langle 0|\varphi^-(\hat{x})\varphi^+(\hat{y})|0\rangle=const$ and
violation of the microcausality principle.

In the relativistic case the entire physical Hilbert space ${\cal H}$ is 
a direct sum of the coherent subspaces where different representations of 
the Poincar\'e group are realized. Roughly speaking, different subspaces 
correspond to different types of particles. Considered in the rest of the paper
are the one-particle states of a free field; we shall restrict ourselves to
the neutral scalar field, spinor field of Dirac electrons and a
gauge field (photons).

Suppose we are given two orthogonal states of the neutral scalar field,
$|\varphi_{1,2}\rangle\in\Omega\subset{\cal H}$
\begin{equation}               
|\varphi_{1,2}\rangle=\varphi^+(f_{1,2})|0\rangle=
\int\varphi^+(\hat{x})f_{1,2}(\hat{x})d\hat{x}|0\rangle,
\end{equation}               
\begin{equation}
\varphi^{\pm}(\hat{x})=\frac{1}{(2\pi)^{3/2}}\int_{V^+_m}
\mbox{e}^{\pm i\hat{p}\hat{x}}a^{\pm}({\bf p})\frac{d{\bf p}}{\sqrt{2p_0}},
\end{equation}               
\begin{displaymath}
|{\bf p}\rangle=a^{+}({\bf p})|0\rangle\in\Omega^*,\quad
\varphi^+(\hat{x})|0\rangle\in\Omega^*,
\end{displaymath}
where $a^{\pm}({\bf p})$ are the field operators in the momentum 
representation which generate the generalized eigenvectors from $\Omega^*$.
The integration in Eq.(16) is performed over the front part of the mass 
shell $V^+_m$ ($p_0^2-{\bf p}^2=m^2$, $p_0>0$). The operator valued
distribution $\varphi^{\pm}(\hat{x})$ satisfies the Klein-Gordon equation
\begin{equation}               
(\Box+m^2)\varphi(\hat{x})=0,\quad 
\varphi(\hat{x})=\varphi^+(\hat{x})+\varphi^-(\hat{x}).
\end{equation}               
Since in the relativistic case, just in the non-relativistic quantum mechanics,
the field states 
$|\varphi_{1,2}\rangle=\varphi^+(f_{1,2})|0\rangle\in \Omega\subset{\cal H}$
are described by the rays in the Hilbert space, the
appropriate measurement has the form analogous to Eq.(1) because
the orthogonal identity resolution of that kind is only based on the 
geometrical properties of ${\cal H}$ (projection on the rays corresponding
to the states $|\varphi_{1,2}\rangle$). The measurement allowing to reliably 
distinguish between the two orthogonal states is given by the identity 
resolution
\begin{equation}               
{\cal P}_1+{\cal P}_2+{\cal P}_{\bot}=I,\quad
{\cal P}_{\bot}=I-{\cal P}_1-{\cal P}_2,
\end{equation}               
\begin{displaymath}
{\cal P}_{1,2}=|\varphi_{1,2}\rangle\langle\varphi_{1,2}|=
\varphi^+(f_{1,2})|0\rangle\langle 0|\varphi^-(f_{1,2}),
\end{displaymath}
\begin{displaymath}
I=\int \varphi^+(\hat{x})|0\rangle\langle 0|\varphi^-(\hat{x}) d{\bf x}=
\int_{V^+_m}|{\bf p}\rangle\langle{\bf p}|\frac{d{\bf p}}{2p_0}.
\end{displaymath}
The above identity resolution is written in terms of the
generalized eigenstates $\varphi^+(\hat{x})|0\rangle\in\Omega^*$.

The probabilities of obtaining the outcomes in different possible channels are
\begin{equation}               
\mbox{Pr}_i\{|\varphi_j\rangle\}=\mbox{Tr}
\{|\varphi_i\rangle\langle\varphi_i|{\cal P}_j\}=
|\langle\varphi_j|\varphi_i\rangle|^2=
|\int\int f^*_j(\hat{x})D^+_m(\hat{x}-\hat{x}')f_i(\hat{x}')d\hat{x}d\hat{x}'|^2=
\end{equation}               
\begin{displaymath}
=|\int f_j^*(\hat{p})f_i(\hat{p})\theta(p_0)\delta(\hat{p}^2-m^2)d\hat{p}|^2=
|\int_{V^+_m}f_j^*({\bf p})f_i({\bf p})\frac{d{\bf p}}{2p_0}|^2=\delta_{ij},
\end{displaymath}
where the generalized commutator function for the field with 
mass $m$ is [23]
\begin{equation}               
D^{\pm}_{m}(\hat{x})=\pm\frac{1}{i(2\pi)^{3/2}}
\int \mbox{e}^{i\hat{p}\hat{x}}\theta(\pm p_0)\delta(\hat{p}^2-m^2) 
d\hat{p}=
\end{equation}               
\begin{displaymath}
\frac{1}{4\pi}\varepsilon({x_0})\delta(\hat{x}^2)\mp
\frac{im}{8\pi\sqrt{\hat{x}^2}}\theta(\hat{x}^2)
\left[N_1(m\sqrt{\hat{x}^2})\mp i\varepsilon(x_0)J_1(m\sqrt{\hat{x}^2})\right]
\pm\frac{im}{4\pi^2\sqrt{-\hat{x}^2}}\theta(-\hat{x}^2)
K_1(m\sqrt{-\hat{x}^2}),
\end{displaymath}
\begin{displaymath}
\varepsilon({x_0})\delta(\hat{x}^2)\equiv
\frac{\delta(x_0-|{\bf x}|) - \delta(x_0+|{\bf x}|)}{2|{\bf x}|}.
\end{displaymath}
To within the exponential tales, the commutator function is only different 
from zero inside the light cone and has a singularity at the cone itself
$\lambda^2=(\hat{x}-\hat{x}')^2=0$; outside the light cone 
the functions $D^{\pm}(\lambda)$ vanish exponentially at the Compton length
as $|\lambda|^{-3/4}\exp{(-m\sqrt{|\lambda|})}$ [12,23,24]. For a fixed point
$\hat{x}$ the contribution to the integral is only given by the points 
$\hat{x}'$ lying inside the light cone issued from the point $\hat{x}$, 
which is actually the consequence of the microcausality principle and 
reflects the impossibility of the faster-than-light field propagation.

The non-zero tails of the commutator function for the massive particles
$m\neq 0$ at the Compton length outside the light cone do not result in
any contradictions with the macroscopic causality [26].

The physical states $\varphi^+(f_i)|0\rangle\in{\cal H}$ corresponding
to two different functions $f_i$ can generally be identical, so that there is
no one-to-one correspondence between the functions  
$f(\hat{x})\in{\cal J}(\hat{x})$ and the states they generate
(the generating functions $f_i$ are only recoverable from the state 
$|\varphi_i\rangle \in {\cal H}$ to within the equivalence class). 
To be more precise, the states are only determined (as it is seen from 
Eqs.(16,17)) by the values of the generating function in the momentum
representation $f(\hat{p})$ on the mass shell $p_0=\sqrt{{\bf p}^2+m^2}$. 
The functions $f_{1,2}({\bf p})$ should be interpreted as the 
restriction of the functions $f_{1,2}(\hat{p})$ 
to the mass shell, so that the states corresponding to any functions
$f(\hat{x})$ coinciding in the $\hat{p}$-representation on the mass shell
are physically identical.

To eliminate this ambiguity it is convenient to rewrite Eq.(20) in
the following equivalent way:
\begin{equation}
\mbox{Pr}_i\{|\varphi_j\rangle\}=\mbox{Tr}
\{|\varphi_i\rangle\langle\varphi_i|{\cal P}_j\}=
|\langle\varphi_j|\varphi_i\rangle|^2=
|\int\int f^*_j({\bf x})D^+_m({\bf x}-{\bf x}')f_i({\bf x}')d{\bf x}d{\bf x}'|^2
=\delta_{ij},
\end{equation}
where
\begin{equation}
f_i({\bf x})=\int \mbox{e}^{-i(t_0p_0({\bf p})-{\bf p}{\bf x})}
f_i({\bf p}) d{\bf p}.
\end{equation}
The commutator function $D^+_m({\bf x}-{\bf x}')$ is obtained from Eq.(21) 
if one sets $t=t'$ in $D^+_m(\hat{x}-\hat{x}')$ ($\hat{x}=(t,{\bf x})$, 
\mbox{ }$\hat{x}'=(t',{\bf x}')$), i.e. the non-zero contributions to the 
measurement at a given time slice $t=t'$ are only given by spatially 
coinciding points. Here $f_i({\bf x})$ is understood as the amplitude
(playing the role similar to the one-particle wavefunction in the position 
representation in the non-relativistic case) taken at all points at the
same moment of time corresponding to the beginning of the measurement
procedure, i.e. to the moment of time starting from which the entire
spatial domain embracing the considered state ($f_i({\bf x})$) 
becomes accessible to the measuring apparatus. 

The measurement defined by Eqs.(20,22) is non-local in the sense
that it requires the access to the entire domain of the coordinate space
${\bf x}$ where the functions $f_i({\bf x})$ are different from zero.
Formally, this measurement can be interpreted as a non-local one in the 
coordinate space performed at a particular moment of time. The device
implementing such a measurement should occupy an extended (formally even 
infinite) domain in the ${\bf x}$-space, i.e. it should have simultaneous
access to the entire state. The non-local measurements of that kind are
not forbidden in the relativistic case. However, the outcome of the measurement
performed with a non-local device cannot be obtained at the same moment as
the measurement starts if the measurement encompasses the entire spatial domain
where the state is present since the information relevant to the measurement 
outcome cannot be gathered instantly to the observer located at a certain 
spatial point from all the points of an extended spatial domain. 
The information can only be gathered in finite time since actually the 
indicated spatial domain should be covered by the past part of the light cone
issued from the point where the observer is located (Fig.1). The entire
domain where the state is present should reside in the interior part of the light 
cone. The minimum time required for the measurement (for fixed input states)
can be determined by examining all the observer positions corresponding to the
situation where the entire domain where the state is present at the time
$t_0$ and which is accessible to the measuring apparatus is completely covered
by the past part of the light cone.
\begin{figure}[tb]
\hspace*{59mm}\special {em:graph fig_1.pcx}
\vspace*{4.6truecm}
\caption{}
\end{figure}

Since the required measurement time is only determined by the condition of
covering the domain by the light cone, it does not depend on the choice of
the reference frame because the light cone is a Lorentz-invariant object.
The latter can be graphically illustrated for a one-dimensional state. 
Suppose that the prepared
state in a certain reference frame has a characteristic extent $x_0-x_1=L$ 
along the $x$-axis. The time required in that system is
$t=L/2c$ (the optimal observer position is at the center of the domain).
At first site, the observer in a moving reference frame would see
a Lorentz-contracted domain where the state is defined. Indeed, 
transformation from the initial reference frame $(x,y,z,t)$ to the moving one
$(x',y',z',t')$ through the hyperbolic rotation (Lorentz transformation)
\begin{equation}
\left(
\begin{array}{c}
x'\\
t'\\
\end{array}
\right)=
\left(
\begin{array}{cc}
\mbox{ch}\psi&\mbox{sh}\psi\\
\mbox{sh}\psi&\mbox{ch}\psi\\
\end{array}
\right)
\left(
\begin{array}{c}
x\\
t\\
\end{array}
\right),
\quad y'=y,\quad z'=z,
\quad \beta=\mbox{th}\psi,
\end{equation}
makes the domain size equal to $x_0'-x_1'=(x_0-x_1)\sqrt{1-\beta^2}$. 
The time necessary for obtaining the information on the state in the
new reference frame is
$t_0'-t_1'=(x_0'-x_1')/2c=(x_0-x_1)\sqrt{1-\beta^2}/2c$ and can be done
arbitrarily small in the moving reference frame. However, the time
elapsed in the initial reference frame
$t_0-t_1=(t_0'-t_1')/\sqrt{1-\beta^2}=(x_0-x_1)/2c$ remains the 
same\footnote{This point is important for construction of cryptographic
protocols, for example, for the quantum cryptography based on orthogonal 
states, since it prohibits eavesdropping employing the twin paradox.}.

For the three-dimensional case the minimal required time will be determined
by the size of the maximum cross-section of the spatial domain.

The functions $f_i({\bf x})$ rapidly decrease at the infinity, 
but do not become identical zero outside of any compact domain actually 
because it is impossible to obtain a function with compact support in the 
position space taking a Fourier transform of a function which is only 
defined on the mass shell in the momentum space (the proof based on the
Wiener--Paley theorem can be found, e.g. in Refs.[27,28]). Formally, this 
means that the time necessary for the reliable identification (with unit
probability) of one of the two orthogonal filed states is infinite since
because of the infinite support of the functions generating the field states
one should have access to the entire coordinate space. However, the states
still can be identified in a finite time (which, of course, depends on the
structure of the chosen states) with the probability arbitrarily close to unity.

Consider now the case of a multicomponent spinor field of Dirac electrons.
The operator valued spinor field distribution has the form [23,24]
\begin{equation}               
\psi(\hat{x})=\frac{\sqrt{m}}{(2\pi)^{3/2}}
\int \frac{d{\bf p}}{\sqrt{\varepsilon_{\bf p}}}
\sum_{\zeta=\pm 1/2}\left\{
a^{(+)}_{\zeta}({\bf p})u^{(+)}_{\zeta}({\bf p})\mbox{e}^{-i\hat{p}\hat{x}}+
a^{*(-)}_{\zeta}({\bf p})u^{(-)}_{\zeta}({\bf p})\mbox{e}^{i\hat{p}\hat{x}}
\right\},
\end{equation}               
and the Dirac conjugate operator is
\begin{equation}               
\tilde{\psi}(\hat{x})=\frac{\sqrt{m}}{(2\pi)^{3/2}}
\int \frac{d{\bf p}}{\sqrt{\varepsilon_{\bf p}}}
\sum_{\zeta=\pm 1/2}\left\{
a^{*(+)}_{\zeta}({\bf p})\tilde{u}^{(+)}_{\zeta}({\bf p})\mbox{e}^{i\hat{p}\hat{x}}+
a^{(-)}_{\zeta}({\bf p})\tilde{u}^{(-)}_{\zeta}({\bf p})\mbox{e}^{-i\hat{p}\hat{x}}
\right\}
\end{equation}               
with the normalization conditions of the Dirac spinors
\begin{equation}               
2\sum_{\zeta=\pm 1/2}
u^{(\pm)\alpha}_{\zeta}({\bf p})\tilde{u}^{(\pm)}_{\zeta\beta}({\bf p})=
\left(\frac{\hat{p}}{m}\right)^{\alpha}_{\beta}\pm\delta^{\alpha}_{\beta},
\quad \varepsilon_{\bf p}=\sqrt{{\bf p}^2+m^2},\quad
\alpha,\beta= 1 \ldots 4.
\end{equation}               
The operator valued distribution (25) satisfies the Dirac equation
\begin{equation}               
(\hat{p}+m)\psi(\hat{x})=0,\quad
\hat{p}=i\gamma^{\mu}\partial_{\mu},
\end{equation}               
and the anticommutation relations
\begin{displaymath}
[\psi^{\alpha}(\hat{x}),\psi^{\beta}(\hat{x}')]_{+}=
[\tilde{\psi}_{\alpha}(\hat{x}),\tilde{\psi}_{\beta}(\hat{x}')]_{+}=0,
\end{displaymath}               
\begin{equation}               
[\psi^{\alpha}(\{\hat{x}),\tilde{\psi}_{\beta}(\hat{x}')]_{+}=-i
S^{\alpha}_{\beta}(\hat{x}-\hat{x}')=
(i\gamma^{\mu}\partial_{\mu}+m)D_m(\hat{x}-\hat{x}'),
\quad D_m(\hat{x})=D_m^+(\hat{x})+D_m^-(\hat{x}).
\end{equation}               
The smeared operator functions can be written in the form [12]
\begin{equation}               
|\psi({\bf f})\rangle=\int \psi^{\alpha}(\hat{x})f^{\alpha}(\hat{x})d\hat{x},
\quad f^{\alpha}(\hat{x})\in {\cal J}(\hat{x}).
\end{equation}               
The two orthogonal states $|\varphi_{1,2}\rangle$  
($\langle\varphi_1|\varphi_2\rangle=0$) of the spinor field can be written
\begin{equation}               
|\varphi_{1,2}\rangle=\left(
\int \psi^{\alpha}(\hat{x})f^{\alpha}(\hat{x})d\hat{x}\right)
|0\rangle \in \Omega\subset {\cal H},
\end{equation}               
with the corresponding identity resolution employing the generalized states
\begin{equation}               
I=\int \left(\psi(\hat{x})\right)|0\rangle
\langle 0|\left(\tilde{\psi}(\hat{x})\right)d{\bf x}=
\sum_{\zeta=\pm\frac{1}{2},s=\pm} 
\int |{\bf p}, \zeta, s\rangle\langle {\bf p},\zeta,s|
\frac{m d{\bf p}}{2\varepsilon({\bf p})},
\end{equation}               
\begin{displaymath}
|{\bf p},\zeta,s\rangle=a^{*(s)}_{\zeta}({\bf p})|0\rangle,
\end{displaymath}
and similarly for the projection operators on the states $|\varphi_1\rangle$ and 
$|\varphi_2\rangle$.

The probabilities of obtaining different measurement outcomes are
\begin{equation}               
\mbox{Pr}_i\{|\varphi_j\rangle\}=\mbox{Tr}
\{|\varphi_i\rangle\langle\varphi_i|{\cal P}_j\}=
|\int\int f^{*\alpha}_{j}(\hat{x})S^{+\alpha}_{m\beta}
(\hat{x}-\hat{x}')f_{i\beta}(\hat{x}')d\hat{x}d\hat{x}'|^2=
\end{equation}               
\begin{displaymath}
|\int f^{*\alpha}_{j}(\hat{p})(\hat{p}-m)^{\alpha}_{\beta}f_{i\beta}(\hat{p})
\theta(p_0)\delta(\hat{p}^2-m^2)d\hat{p}|^2=
|\int_{V^+_m} f^{*\alpha}_{j}({\bf p})(\hat{p}-m)^{\alpha}_{\beta}
f_{i\beta}({\bf p})\frac{d{\bf p}}{2\varepsilon({\bf p})}|^2=
\delta_{ij},
\end{displaymath}
Here $f_{i\beta,j\alpha}({\bf p})$ are the values of the amplitude on the mass
shell with positive energy. In the present case, just as for the scalar 
field, the answer is expressed through the derivative of the commutator
function $D^{\pm}(\hat{x}-\hat{x}')$ so that everything said above on the
finiteness of the time required for the distinguishing between two states
applies as well to the fermionic field.

Consider now the case of a gauge field which is most interesting from the 
viewpoint of applications, i.e. the photon field. The electromagnetic field
operators are written as [12]
\begin{equation}               
A^{\pm}_{\mu}(\hat{x})=\frac{1}{(2\pi)^{3/2}}\int\frac{d{\bf k}}{\sqrt{2k_0}} 
\mbox{e}^{\pm i \hat{k}\hat{x}} e_{\mu}^{m}({\bf k})a^{\pm}_{m}({\bf k}) 
\end{equation} 
and satisfy the commutation relations 
\begin{equation} 
[A^{-}_{\mu}(\hat{x}),A^{-}_{\nu}(\hat{x}')]_{-}=ig_{\mu\nu}
D_{0}^{-}(\hat{x}-\hat{x}'),
\end{equation} 
where $D_{0}^{-}(\hat{x}-\hat{x}')$ is the commutator function for the
massless field (21). There are four types of photons:  two transverse,  
one longitudinal, and one temporal. The two latter types are actually 
fictitious particles and can be eliminated at the expense of introducing
an indefinite metric [12]. For our purposes the shortest way to the required 
result consists in employing a specific gauge. We shall further work in the 
subspace of physical space using the Coulomb gauge 
$A_{\mu}=({\bf A},\varphi=0)$ dealing with the two physical transverse
states of the electromagnetic field.

The operator valued distribution is a vector in the three-dimensional space:
\begin{equation} 
\vec{\mbox{\boldmath $\psi$}}(\hat{x})=\frac{1}{(2\pi)^{3/2}}
\int_{V^+_0}\frac{d{\bf k}}{\sqrt{2k_0}}
\sum_{s=\pm 1}{\bf w}({\bf k},s)\{a({\bf k},s)\mbox{e}^{-i\hat{k}\hat{x}}+
a^+({\bf k},-s)\mbox{e}^{i\hat{k}\hat{x}}\};
\end{equation} 
here ${\bf w}({\bf k},s)$ is a three-dimensional vector describing the
polarization state $s=\pm 1$,
\begin{equation} 
{\bf w}({\bf k},\pm)=\frac{1}{\sqrt{2}}
[{\bf e}_1({\bf k}) \pm i{\bf e}_2({\bf k})],\quad
{\bf e}_1({\bf k})\bot {\bf e}_2({\bf k}),\quad
|{\bf w}({\bf k},s)|^2=1,
\end{equation} 
where ${\bf e}_{1,2}({\bf k})$ are the orthogonal vectors normal to ${\bf k}$. 
The field operator satisfies the Maxwell equations
\begin{equation} 
\nabla\times\vec{\mbox{\boldmath $\psi$}}(\hat{x})=
-i\frac{\partial}{\partial t}\vec{\mbox{\boldmath $\psi$}}(\hat{x}),
\end{equation} 
\begin{displaymath}
\nabla\cdot\vec{\mbox{\boldmath $\psi$}}(\hat{x})=0.
\end{displaymath}               
The smeared field operators can be written as
\begin{equation} 
\vec{\mbox{\boldmath $\psi$}}(f_{1,2})=
\sum_{s=\pm 1}\int\vec{\mbox{\boldmath $\psi$}}(\hat{x},s)
f_{1,2}(\hat{x},s)d\hat{x}=
\end{equation} 
\begin{displaymath}
\frac{1}{(2\pi)^{3/2}}
\int_{V^+_0}\frac{d{\bf k}}{\sqrt{2k_0}}
\sum_{s=\pm 1}{\bf w}({\bf k},s)\{
f_{1,2}({\bf k},s) a({\bf k},s)\mbox{e}^{-i\hat{k}\hat{x}}+
f_{1,2}({\bf k},-s) a^+({\bf k},-s)\mbox{e}^{i\hat{k}\hat{x}}\},
\end{displaymath}               
where the values of the functions $f_{1,2}({\bf k},s)$ are taken at the mass
shell (light cone $V^+_0$ in the momentum representation).
Two orthogonal states of the photon field can be written in the form
\begin{equation} 
\vec{|\mbox{\boldmath $\psi$}}_{1,2}\rangle=\left(
\vec{\mbox{\boldmath $\psi$}}(f_{1,2})\right)|0\rangle.
\end{equation}
The corresponding orthogonal projectors on the states 
$|\vec{\mbox{\boldmath $\psi$}}_{1,2}\rangle$ are
\begin{equation}
\mbox{\boldmath ${\cal P}$}_{1,2}=
|\vec{\mbox{\boldmath $\psi$}}_{1,2}\rangle
\langle\vec{\mbox{\boldmath $\psi$}}_{1,2}|,
\end{equation}
\begin{equation}
I=\int \left(
\vec{\mbox{\boldmath $\psi$}}^+(\hat{x})\right)|0\rangle
\langle 0|\left(\vec{\mbox{\boldmath $\psi$}}^-(\hat{x})\right)
d{\bf x}=
\sum_{s=\pm 1}\int_{V^+_0} 
\left( {\bf w}({\bf k},s)a^+({\bf k},s)\right)
|0\rangle\langle 0|
\left( {\bf w}({\bf k},s)a({\bf k},s)\right)
\frac{d {\bf k}}{2|{\bf k}|}.
\end{equation}
The probabilities of obtaining different measurement outcomes are
\begin{equation}
\mbox{Pr}_i\{|\vec{\mbox{\boldmath $\psi$}}_j\rangle\}=\mbox{Tr}
\{|\vec{\mbox{\boldmath $\psi$}}_i\rangle
\langle\vec{\mbox{\boldmath $\psi$}}_i|\mbox{\boldmath ${\cal P}$}_j\}=
|\langle \vec{\mbox{\boldmath $\psi$}}_j|
\vec{\mbox{\boldmath $\psi$}}_i\rangle|^2=
\end{equation}
\begin{displaymath}
|\int\int f^*_{j}({\bf x})D^{+}_{0}({\bf x}-{\bf x}')f_{i}
({\bf x}')d{\bf x}d{\bf x}'|^2=
|\int_{V^+_0}f^{*}_{i}({\bf k})f_j({\bf k})
\frac{d{\bf k}}{2|{\bf k}|}|^2=\delta_{ij},
\end{displaymath}
where $D^{+}_{0}({\bf x}-{\bf x}')$ is the commutator function for the 
massless field at the time slice $t=t'$ has the form 
\begin{displaymath}
D^{+}_{0}({\bf x}-{\bf x}')=
-\frac{1}{4\pi}\frac{\delta(|{\bf x}-{\bf x}'|)}{2|{\bf x}-{\bf x}'|}.
\end{displaymath}

It should be noted that arising in the non-relativistic case instead
of the commutator function $D^+_0$ is the usual 
$\delta({\bf x}-{\bf x}')$-function (see Eq.(12)). The latter is actually
related to the fact that in the non-relativistic case the integration
in the scalar product in the momentum representation is performed with
the Galilean-invariant measure $d\mu({\bf p})=d{\bf p}$ instead of the
Lorentz-invariant measure 
$d\mu(\hat{p})=\theta({p_0})\delta(\hat{p}^2)d\hat{p}$ 
residing at the mass shell in the relativistic case. 

The time necessary for the reliable identification of one of the
pair of orthogonal states is determined by the spatial localization
of the generating functions $f_{1,2}({\bf k})$ in the coordinate
(position) representation or, to be more precise, by the localization of
$|{\bf f}_{1,2}({\bf x},t,s)|^2$
$|{\bf f}_{1,2}({\bf x},t,s)|^2$
\begin{equation}
{\bf f}_{1,2}({\bf x},t,s)=\int_{V^+_0}{\bf w}({\bf k},s)f_{1,2}({\bf k},s)
\mbox{e}^{-i(|{\bf k}|t-{\bf kx})}  d{\bf k},
\end{equation}
where the integration is performed over the mass shell $V^+_0$ (surface
of the light cone in the momentum representation). It has long been known
(e.g., see Ref.[29] and references therein) that it is impossible to obtain
a strictly localizable (with compact support) function in the coordinate space
from a normalized function defined on the mass shell. The latter holds for
both massive and massless particles. However, there exist 
the functions with arbitrarily close to the exponential fall off
at the infinity. Recently, the one-particle photon states of that kind
were explicitly constructed [27]. The restrictions on the arbitrarily close
to the exponential fall off stem from the Wiener-Paley theorem for
square integrable functions [29] and the impossibility of mixing of the
positive and negative frequency states on the mass shell for one-particle 
states.

The possibility of the existence of the states whose localization is
arbitrarily close to the exponential one is actually related to the choice 
of the test function space ${\cal J}(\hat{x})$ which contains the dense
subspace of compactly supported functions ${\cal D}(\hat{x})$. The latter
circumstance is closely related to the local nature of the theory since
the existence of a dense set of compactly supported functions defined
in the Minkowski $\hat{x}$-space allows to achieve the local properties
of the distributions, including the commutator functions, appearing in
the microcausality principle.

Reduction of the set of the test functions can result in the non-local nature
of the theory (e.g., see Refs.[23,29--32]). However, non-locality of the theory
does not imply violation of the causality principle at the
macroscopic level (at the level of the observer) [29,32].

\section{Conclusion}
Thus, the reliable (with unit probability) identification of one-particle
field states generally requires an infinite time because of the requirement
of having access to the entire spatial domain where the state ``is present''.
Formally, the construction of the measurement at the level of the
corresponding identity resolution in the relativistic case is completely
analogous to the non-relativistic quantum mechanics since in both cases the 
states are described by the rays in the Hilbert state. At this level, 
the description of measurement involves only the geometrical properties
of the state space (the scalar product and projections on the corresponding
orthogonal states). The difference between the non-relativistic and relativistic 
cases arises at the level of the internal structure of the scalar product
and conveying the measurement outcome to the observer. In the non-relativistic
case, there are no restrictions on the instantaneous spatially non-local 
measurements and conveying their outcomes to the observer (which can be located 
at an arbitrary point) since there are no limits on the maximum speed of 
information transfer. In the relativistic case the Lorentz-invariant
scalar product already contains (through the commutator functions reflecting
the microcausality principle) the information on the maximum speed of 
the field propagation.  The latter should only be interpreted as the
impossibility for the field to propagate into an extended device (fill it)
faster than light. A similar situation occurs if an extended device moves into 
the domain where the field is present. However, the presence of the field 
in the device alone does not yet mean obtaining of the measurement outcome.
The measurement outcome should be conveyed to a single point (location of the 
observer). The latter can only be done in a finite time depending on the
spatial localization of the states since the spatial domain should be 
``covered'' by the past part of the light cone issued from the point
where the observer is situated (the observer can be interpreted as a classical
device registering the final measurement outcome).

In the above approach all the information on the non-local (extended) device
is contained in the projectors on the corresponding rays in the Hilbert 
space. Formally, these projectors should be considered as an (non-local)
observable which, from the viewpoint of the measurement theory, is quite 
similar to a non-local in the coordinate space observable (e.g., momentum).
Therefore, associated with this observable should be a physical apparatus
implementing the corresponding measurement.

The conclusion on the finiteness of the time interval required for
the state identification does not depend on a particular measurement 
procedure. It is often experimentally easier to realize a non-local 
measurement as a local one employing a localized auxiliary system $A$
interacting with the measured system (it is known that any measurement
can be realized in this way [18,19]). In the course of the joint evolution
the state of the auxiliary system is gradually changed and finally the 
measurement is performed over that system. For the orthogonal states
there are no restrictions on the existence of non-disturbing measurements
[3] (see Eqs.(1--3)) in the sense that just after the measurement
only the state of the auxiliary system $A$ is changed while state of the 
studied system $S$ coincides with its initial state just before the
interaction between $S$ and $A$ was turned on. In that case it is also
intuitively clear that a finite time is required (if the field state is 
extended) for the field to pass through a point where the system $A$
is localized and the interaction between $A$ and $S$ occurs which cannot
be done faster than $t\sim L/c$ because of the existence of a maximum speed 
of the field propagation. It is also clear that this time cannot be reduced
by enhancement of the local coupling between $A$ and $S$ because this
time is limited by the field propagation speed. The required time cannot
also be reduced employing the Lorentz contraction of the domain where
the field is present occurring in a moving reference frame 
(see discussion above). Naturally, the auxiliary system $A$ should also be
described by the quantum field theory and one should assume that it is 
so strongly spatially localized that the time necessary to perform a 
measurement on this system can be done arbitrarily small. However,
it is much more difficult to obtain any general results in this approach
since it inevitably requires making certain assumptions concerning
the system $S$ itself and its interaction with the auxiliary system $A$.

Formally, the von Neumann measurement described by the orthogonal resolution 
of identity given by Eq.(43) in the subspace of one-particle states in the 
Hilbert space does not disturb the orthogonal states since the situation
here is analogous to the non-relativistic case described by Eqs.(1--4).
However, because of the non-local nature of the projectors, the reliable
and non-disturbing identification of the orthogonal states achieved by
this measurement formally requires infinite time.

With respect to the time necessary for the measurement, the situation is quite
similar for the non-orthogonal states, the only difference being that
the latter cannot be reliably distinguished even in an infinite time.

In conclusion, it should be noted that the non-relativistic quantum
cryptographic protocols implicitly assume the possibility of performing
a measurement distinguishing between the two states in a finite time
(ideally, an instantaneous measurement). Briefly, all the non-relativistic
protocols can be described in the following way. User A sends either the state
$|\psi_1\rangle$ or  $|\psi_2\rangle$ to user B who performs
quantum-mechanical measurements. Part of information on the measurement
outcomes is discussed through the open communication channel. It is assumed
that the states $|\psi_1\rangle$ or $|\psi_2\rangle$ are prepared and
measured in certain agreed moments of time so that the preparation and
measurement procedures can be performed at arbitrary moments of time.
Thus, it is implicitly assumed that the protocol can be implemented in a 
finite time interval.

The relativistic quantum field theory does not allow access to the
entire state in any finite time because of the non-localizable nature
of the state. Therefore, all the exchange protocols with finite duration
should inevitably involve an error in the preparation and detection
of both orthogonal and non-orthogonal one-particle states even for the
ideal communication channel. In other words, the non-localizability
of the states results in the effective ``noise'' because of the impossibility
of the reliable detection of orthogonal states in a finite time
although for a fixed measurement duration the error can be made arbitrarily 
(exponentially) small by choosing more and more strongly localized state.
This circumstance should be taken into account in the realistic
cryptographic protocols employing photons as the information carriers.

The authors are grateful to W.L.Golo, S.V.Iordanskii and V.M.Edelstein for
the interest and discussion of the results of this work. 

This work was supported by the Russian Foundation for Basic Research
(project No 99-02-18127), the project ``Physical Principles of the
Quantum Computer'',  and the program ``Advanced Devices and Technologies
in Micro- and Nanoelectronics'' (project No 02.04.5.2.40.T.50).

This work was also supported by the Wihuri Foundation, Finland.

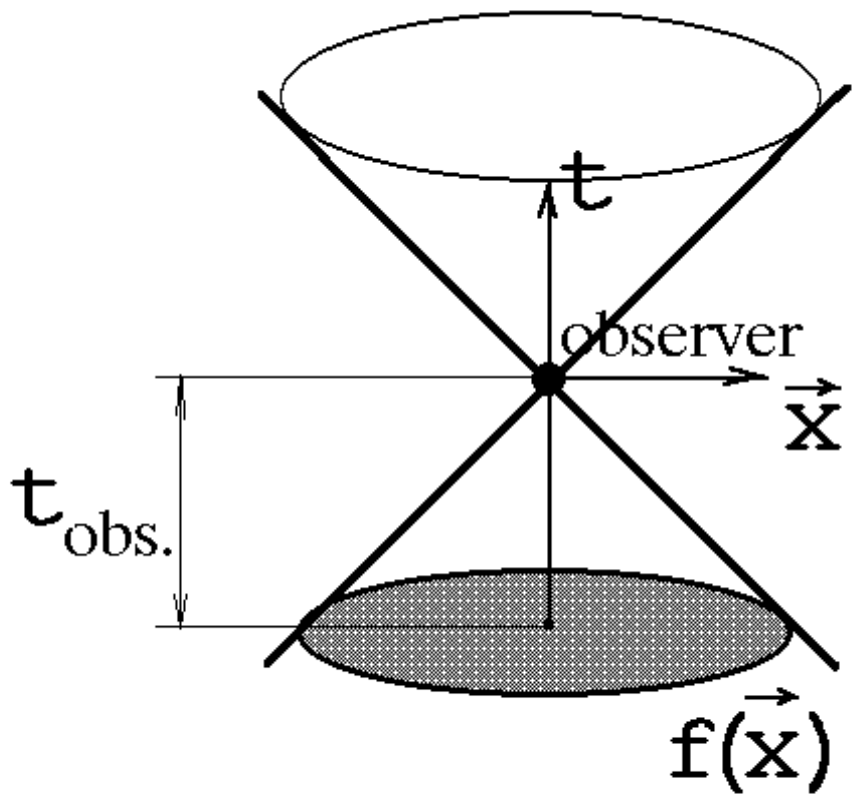
\end{document}